\documentclass{iaus}
\usepackage{graphicx}

\title[Boron Abundances in the Galactic Disk] %% give here short title %%
{Boron Abundances in the Galactic Disk}

\author[Katia Cunha]   %% give here short author list %%
{Katia Cunha$^1$}
\affiliation{$^1$NOAO \\ 
950 N. Cherry Ave, Tucson , Arizona \\ email: {\tt kcunha@noao.edu}}

\pubyear{2010}
\volume{268}  %% insert here IAU Symposium No.
\jname{Light elements in the Universe}
\editors{C. Charbonnel, M. Tosi, F. Primas \& C. Chiappini, eds.}
\begin{document}

\maketitle

\begin{abstract}
When compared to lithium and beryllium, the absence of boron lines in the optical 
results in a relatively small data set of boron abundances measured in Galactic stars to date.
In this paper we discuss boron abundances published in the literature and focus on
the evolution of boron in the Galaxy as measured from pristine boron abundances 
in cool stars as well as early-type stars in the Galactic disk.
The trend of B with Fe obtained from cool F-G dwarfs in the disk is found to have a slope of ~0.87 $\pm$ 0.08
(in a log-log plot).
This slope is similar to the slope of B with Fe found for the metal poor halo stars and there seems to be
a smooth connection between the halo and disk in the chemical evolution of boron.
The disk trend of boron with oxygen has a steeper slope of ~1.5. This slope suggests an intermediate behavior between 
primary and secondary production of boron with respect to oxygen.
The slope derived for oxygen is consistent with the slope obtained for Fe provided that [O/Fe] 
increases as [Fe/H] decreases, as observed in the disk.
\end{abstract}

\firstsection % if your document starts with a section,
\section{Introduction}

The light element boron is one of the few elements whose production is not dominated by nucleosynthesis 
in stars, nor by nucleosynthesis occurring in the Big Bang. In fact, it has been known now for almost 
4 decades that Galactic Cosmic Rays are related to the formation of the light elements
and, in particular, of boron (Reeves, Fowler \& Hoyle 1970). The connection between boron production and
cosmic rays spallation reactions, which involve C, N, O atoms, as well as protons and $\alpha$ particles, 
makes boron an interesting element whose abundance evolution in the Galaxy probes the history 
of cosmic rays in the galactic environment. In addition, boron is also proposed to be produced by
neutrino nucleosynthesis occurring in core collapse of Supernovae Type II (Woosley et al. 1990).   

Unveiling the underlying behavior of boron with metallicity (iron and oxygen abundances) 
is crucial in order constrain models for boron production.
One of the challenges in trying to pin down the behavior of boron with metallicity, however, comes
first from the fact that boron is fragile and easily destroyed in stellar interiors
(although sturdier than Li and Be) and, in addition, from the difficulty in obtaining boron observations. 
In this paper we discuss stellar boron abundance results mainly for disk stars which have
been published in the literature. 
Unfortunately, no new boron observations were available in recent years due to the failure of STIS on board HST. 

\section{Boron Transitions and Abundance Determinations}

Boron abundance results are still sparse as boron abundances can only be measured 
from transitions which fall mainly in the ultraviolet.
Boron abundance indicators in different temperature regimes are from three ionization stages: 
neutral boron in cool stars (B I at 2497.723\AA); B II (at 1362\AA) mostly in A-type stars and 
the B III resonance doublet (at 2060\AA) in B-type stars. 

\subsection{Boron in Cool Stars and the Sun}

Boron was measured in the Sun by Kohl, Parkinson \& Withbroe (1977).
One of the pioneering studies of boron abundances in stars was by Boesgaard \& Heacox (1978).  
A few studies appeared more than a decade later from observations obtained with the Hubble Space Telescope
(Duncan, Lambert, \& Lemke 1992; Duncan et al. 1997; Primas et al. 1999). 
The sample analyzed in Duncan et al. (1997) was mostly for
halo stars. Their results were particularly important as 
they found that boron abundances scaled linearly (in log-log space) with the abundance of metals, 
in contrast with predictions from the standard models of cosmic ray production, which proposed a secondary
behavior for boron with metallicity. These predictions from cosmic ray models had remained unchallenged 
for $\sim$ 20 years.  

\begin{figure}[b]
% \vspace*{-2.0 cm}
\begin{center}
\includegraphics[width=2.9in]{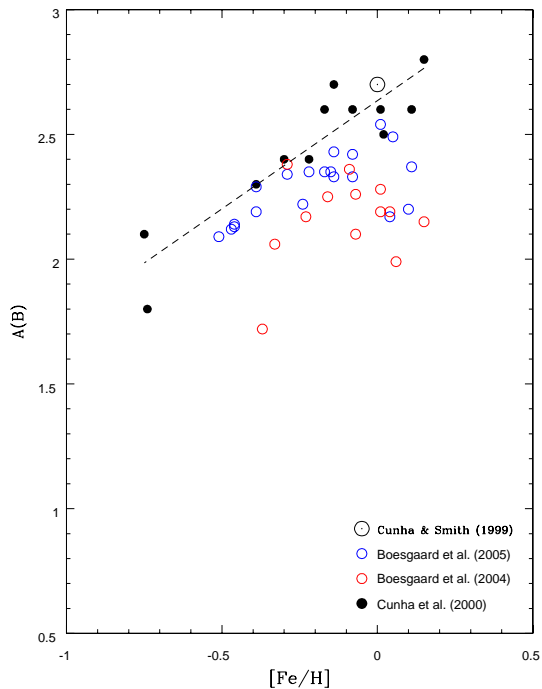} 
%\includegraphics[width=2.6in]{bvsfedisk2.ps}
% \vspace*{-1.0 cm}
\caption{Boron results for the available samples of F-G disk stars. The solar
abundance is also plotted. The targets in Boesgaard et al. (2004) and 
Cunha et al. (2000) are beryllium undepleted, while the stars analyzed in Boesgaard et al. (2005)
are mixed showing a degree of Be depletion. For the Be undepleted stars it is clear that there is a systematic 
abundance offset between the studies of Cunha et al. (2000) and Boesgaard et al. (2004) which
can be attributed to the differences in the line lists adopted in the analyses.
The slope representing the trend of boron with iron for the disk is 0.87 $\pm$ 0.08
(dashed line); this defines the upper envelope (undepleted) of the distribution.}
   \label{fig1}
\end{center}
\end{figure}

Following studies focused on samples of stars with disk metallicities and, by the nature
of their sample, these probed the behavior of boron in the most metal rich stars in the Galaxy
(Boesgaard et al. 1998; Cunha \& Smith 1999; Cunha et al. 2000; Boesgaard et al. 2004; Boesgaard et al. 2005). 
It is important to acknowledge, however, that the line list in the spectral region of the B I 
transition is a major challenge for the analysis of boron in solar-like stars as the spectral region 
to be synthesized is covered with 
strong blending lines for which, in many instances, there is no atomic data available. 
Note, however, that this does not represent a severe problem for the analysis of halo stars as 
the metal lines are vanishingly weak. For solar metallicity stars,
different studies in the literature constructed and adopted different line lists 
which resulted in systematic differences in the derived boron abundances. In the following
we briefly present some of the boron results in these studies.

Cunha et al. (2000) analyzed dwarf stars with [Fe/H] $>$ -1.0 (T$_{eff}$'s between 5650 - 6700K) 
from HST archival data. One important aspect of that study in comparison with Boesgaard et al. (1998; 2004) is 
that the line list 
adopted in the calculation of model spectra in Cunha et al. (2000) was empirically adjusted in order to 
fit the Sun. In using the Sun as a benchmark in the study of solar-type stars,
Cunha \& Smith (1999) re-visited the analysis of boron in the solar
photosphere. In particular, significant effort was put in that study into evaluating and updating the opacities 
which are important in the ultraviolet and which affect the derived boron abundances.
This resulted in the revision upwards of the boron abundance in the solar photosphere,
which was found to be in good agreement with the boron meteoritic value of A(B)= 2.79 $\pm$ 0.04 (
see Lodders, Palme \& Gail 2009). 
The agreement between the boron abundances in the solar photosphere and meteorites 
(implying an $\it{absence}$ of boron depletion in the Sun) is an important result as 
it reconciles with the most recent assessment of the beryllium abundance in the solar photosphere by
Asplund et al. (2009), indicating no beryllium depletion in the Sun. If Be is indeed not 
depleted in the solar photosphere, it follows that boron, which is less fragile than Be, cannot be 
depleted in the solar photosphere.

Boron abundances for disk dwarfs with effective temperatures close to solar are shown in Figure 1. 
Non-LTE corrections for the B I transition at 2497\AA\ for this temperature range at
solar metallicity are deemed to be small (Kiselman \& Carlson 1996).
In order to have all stars and the Sun on a consistent scale, all disk stars in Cunha et al. (2000)
were analyzed homogeneously.  The study by Boesgaard et al. (2004) used a different line list which
was not fine tuned in order to fit the solar spectrum.
It can be seen that the results from Cunha et al. (2000) and Boesgaard et al. (2004), 
all for targets with undepleted beryllium, have a small systematic offset. It is clear also that 
the targets analyzed Boesgaard et al. (2005) have significantly lower boron abundances,
but this is expected as the target stars were selected in that study to be beryllium depleted in order
to further study mixing.

%Boesgaard et al. (1998) > 9 dwarfs; [Fe/H] from -0.75 to +0.15
%Cunha & Smith (1999) : Re-Analysis of the Sun : an effort to place solar type near-metallicity stars on one consistent scale
%Cunha et al. (2000) > 14 FG-dwarfs from HST archival data  > most of them beryllium undepleted 
%Boesgaard et al. (2004) > 16 dwarfs with undepleted beryllium 
%Boesgaard et al. (2005) > 13 beryllium depleted stars; study mixing 

\subsection{Boron in Early-type Stars}

Of the light element trio, boron is the only element whose abundance can be measured in early-type stars. 
One problem in using early-type stars to define the boron Galactic trend, however, is the varying amounts 
of boron depletion in OB-type stars, as depletion of boron is proportional to stellar mass, age and rotational velocity.
In addition, unlike the case of observations of Li and Be in cool stars, there is not a sensitive monitor 
of depletion in early-type stars. 
Boron is burnt at temperatures which are lower than those at which the CN cycle takes place and is
much more sensitive to mixing than nitrogen.

% Venn, Lambert & Lemke (1996) : Sample of 6 B-type stars (more evolved) ; 5 upper limits [B II from IUE SWP]
%    2)  Cunha et al. (1997): Sample of  4 Orion  B-stars [B II GHRS]
%    4) Venn et al. (2002) : B III [HST/STIS] in 7 MS stars; 4 boron upper limits
%    5) Mendel et al. (2006): B III [HST/STIS] in 7MS stars; most of them with  depleted boron; 1 boron upper limit

Some studies in the literature have derived boron abundances from HST observations obtained with the GHRS and STIS spectrographs
in relatively small samples of early-type stars (Cunha et al. 1997; Venn et al. 2002; Mendel et al. 2006).
Most of these boron abundances, however, were found to be somewhat mixed and therefore not representative of the 
chemical composition of the gas which formed these young stars. The larger sample analyzed by Proffitt \& Quigley (2001) 
from IUE archival observations of the B III resonance line at 2066\AA, although not having the same 
spectral quality as HST data, found some stars to be boron undepleted which helped define the disk trend.

\section{Boron Abundance Trends in the Disk}

\begin{figure}[b]
% \vspace*{-2.0 cm}
\begin{center}
\includegraphics[height=0.92\textwidth,angle=-90]{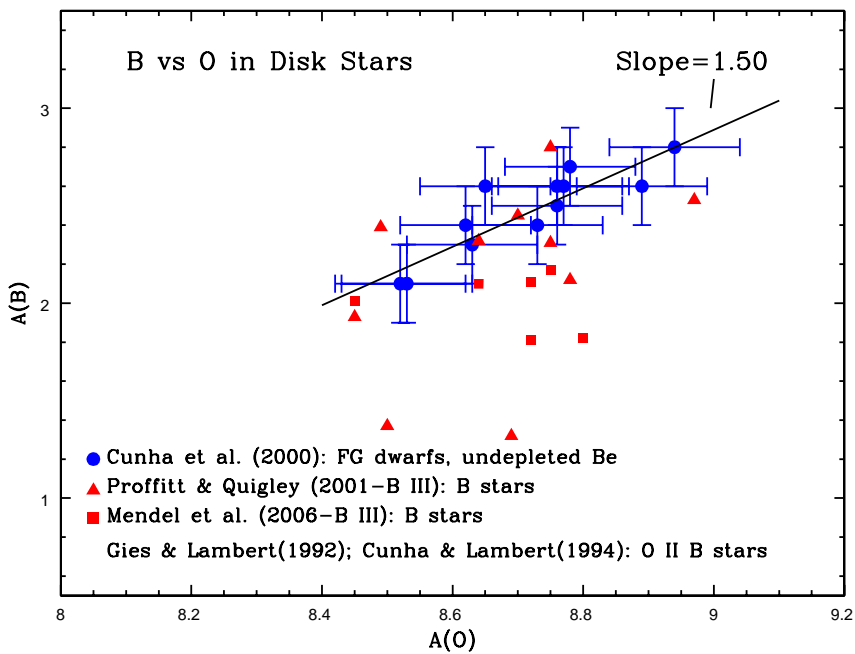}
%\includegraphics[width=3.2in]{bo.fgb2.ps}
% \vspace*{-1.0 cm}
 \caption{The evolution of boron and oxygen abundances in the Galactic disk.
Boron results for cool disk FG-type dwarfs are taken from Cunha et al. (2000); these
targets have undepleted Be abundances which indicate that their boron content is
not mixed and representative of their natal clouds. The evolution of boron and
oxygen from this dataset can be represented by linear relation with slope $\sim$1.5. Boron results for
early-type stars from Proffitt \& Quigley (2001) and Mendel et al. (2006) are also
shown. The lower boron abundances in some of the B-type stars are due to internal mixing
and astration. 
} 
   \label{fig1}
\end{center}
\end{figure}
 
The evolution of boron with oxygen for metallicities covering the range spanned by the disk
is shown in Figure 2. The blue filled circles represent the
FG-dwarfs analyzed in Cunha et al. (2000) with the errorbars indicating the estimated
abundance uncertainties. 
Boron results for early-type stars from Proffitt et al. (2001; filled red triangles) 
and Mendel et al. (2006; filled red squares) are also shown. 
Most of the B stars shown have roughly undepleted boron and on average follow the
disk trend delineated by the cool stars. The general agreement between the results in cool and
hot stars is pleasing given that the physical conditions in their stellar atmospheres
are quite distinct.
The behavior of boron with oxygen 
can be represented by a linear relation (in the log-log plot) with a slope $\sim$1.5, 
which can suggest an intermediate behavior between primary and secondary production for boron with respect to oxygen.

\begin{figure}[t]
% \vspace*{-2.0 cm}
\begin{center}
\includegraphics[height=0.92\textwidth,angle=-90]{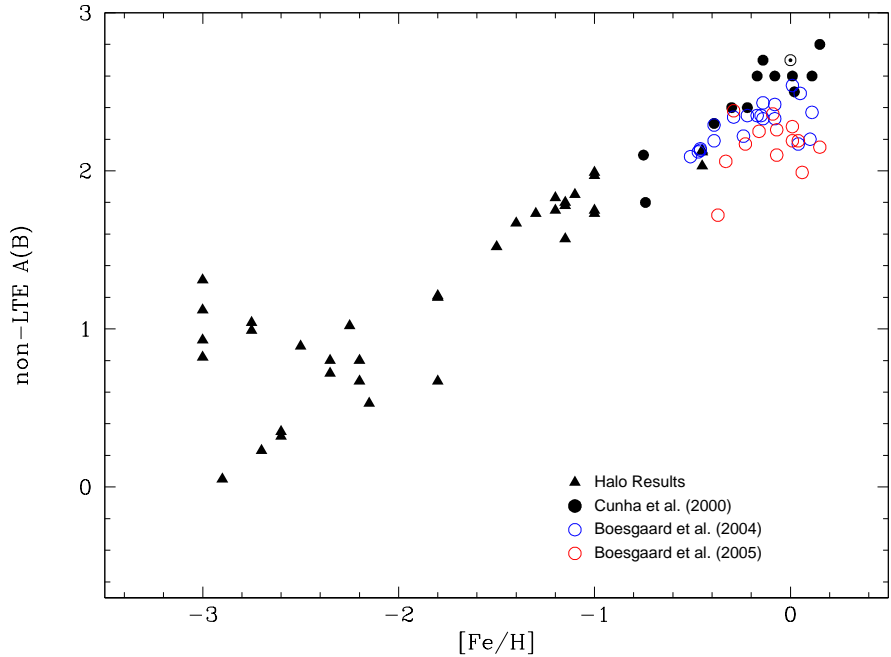}
%\includegraphics[width=3.2in]{bo.fgb2.ps}
% \vspace*{-1.0 cm}
 \caption{
The behavior of boron with metallicity for the Galactic disk in comparison 
with the trend obtained for the more metal poor stars in the halo.
If the abundance results in Cunha et al. (2000) are adopted as representative of the disk (filled circles), 
there seems to be a smooth transition between the disk and halo. 
The halo star abundances are taken from Duncan et al. (1997) and Garcia Lopez et al. (1998).}
   \label{fig1}
\end{center}
\end{figure}

In Figure 3 we plot boron abundances versus [Fe/H] spanning the metallicity range from the halo to the disk. 
If we adopt the boron abundances for the Be undepleted stars from Cunha et al. (2000; filled circles) 
as representative of the disk value, there seems to be a smooth transition between the halo and disk
which follows a slope of $\sim$ 0.9. This slope for the disk + halo is closer to a primary rather than a
secondary behavior for boron production. In addition, it is important to note that
the slope derived for oxygen (from Figure 2) is consistent with the one obtained for Fe provided 
that [O/Fe] increases as [Fe/H] decreases, as observed in the disk.

\section{How Do We Move Forward?}

Probing the behavior of boron with metallicity is crucial for understanding boron production in the Galaxy
and the relatively small number of stars analyzed to date could usefully be increased.
It is good news that the STIS spectrograph has been recently fixed in a very successful NASA servicing mission (SM4)
and that it can now be ready for more boron observations in the UV. In addition, the new UV Cosmic
Origins Spectrograph (COS) was deployed on HST during SM4.
We also need improvements in the abundance analysis and in particular improvements in the line lists to model the 
boron region. Full non-LTE treatment is needed, including non-LTE analysis of transitions of all
elements contributing to the boron blend; and a complete hydrodynamic 3-D modelling of the
stellar atmospheres is also something to look forward to in particular for the cool stars.

\end{document}